\documentclass[journal,comsoc]{IEEEtran}
\usepackage{amsfonts}
\usepackage[T1]{fontenc}

\ifCLASSINFOpdf
\else
\fi
\usepackage[pagewise]{lineno}

\usepackage{amsmath}
\usepackage{epsfig}
\usepackage{graphicx}
\usepackage{psfig}
\usepackage{times}
\usepackage{CJK}
\usepackage{makecell}
\usepackage{epsf}
\usepackage{bm} 
\usepackage{slashbox}
\usepackage{amsthm}
\usepackage{cite}
\usepackage{booktabs}
\usepackage{color}
\usepackage{float}
\usepackage{subfigure}
\usepackage{graphicx}

\usepackage{array}
\begin{document}

\title{Multi-User Hybrid Precoding for Dynamic Subarrays in MmWave Massive MIMO Systems}
%
%
%
\author{\IEEEauthorblockN{ Jing Jiang,~\IEEEmembership{Member,~IEEE,} Yue Yuan, and Li Zhen,~\IEEEmembership{Member,~IEEE}}

\thanks{Jing Jiang, Yue Yuan and Li Zhen are with the Shaanxi Key Laboratory of Information Communication Network and Security, Xi'an University of Posts and Telecommunications, Xi'an, China e-mail: YuanYue12Artemis@163.com. This work has been submitted to the IEEE Journal for possible publication. Copyright may be transferred without notice, after which this version may no longer be accessible.}}

%
%

\markboth{Journal of \LaTeX\ Class Files,~Vol.~14, No.~8, January~2019}%
{Shell \MakeLowercase{\textit{et al.}}: Multi-User Hybrid Precoding for Dynamic Subarrays in MmWave Massive MIMO Systems}
%



\maketitle

\begin{abstract}
Dynamic subarray achieves a compromise between sum rate and hardware complexity for millimeter wave (mmWave) massive multiple-input multiple-output (MIMO) systems in which antenna elements are dynamically partitioned to radio frequency (RF) chain according to the channel state information. However, multi-user hybrid precoding for the dynamic subarray is intractable to solve as the antenna partitioning would result in the user unfairness and multi-user interference (MUI). In this paper, a novel multi-user hybrid precoding framework is proposed for the dynamic subarray architecture. Different from the existing schemes, the base station (BS) firstly selects the multi-user set based on the analog effective channel. And then the antenna partitioning algorithm allocates each antenna element to RF chain according to the maximal increment of the signal to the interference noise ratio (SINR). Finally, the hybrid precoding is optimized for the dynamic subarray architecture. By calculating SINRs on the analog effective channels of the selected users, the antenna partitioning can greatly reduce computation complexity and the size of the search space. Moreover, it also guarantees the user fairness since each antenna element is allocated to acquire the maximal SINR increment of all selected users. Extensive simulation results demonstrate that both the energy efficiency and sum rate of the proposed solution obviously outperforms that of the fixed subarrays, and obtains higher energy efficiency with slight loss of sum rate compared with the fully-connected architecture.
\end{abstract}


%
\IEEEpeerreviewmaketitle

\section{Introduction}
%
%
%
%
\IEEEPARstart{M}{illimeter} wave (mmWave) systems have been identified as a promising solution to cope with the explosive growth of mobile traffic [1], [2]. Moreover, massive multiple-input multiple-output (MIMO) can provide significant array gains to compensate for the serve propagation losses and improve the system capacity of mmWave systems. For mmWave Massive MIMO systems, hybrid precoding is efficient transceivers which can achieve performance close to that of a fully digital precoding with limited number of RF chains [3]-[5].

As is known, fully-connected architecture is widely adopted in hybrid precoding systems [5]-[8]. In this architecture, each RF chain is connected to all the antennas with phase shifters (PSs) and RF adders. Therefore, both the hardware cost and power consumption are high since the numbers of the PSs and RF adders increase linearly with the number of antennas. To address these challenges, it has drawn the tremendous attention on the mapping methodologies of the radio frequency (RF) chains and antennas to reduce the number of PSs and RF adders [9], [10].

Compared with the fully-connected architecture, the partial-connected architecture can greatly reduce the number of PSs and eliminate the need for RF adders, which has low hardware cost and power consumption [9]-[12]. The partial-connected architecture can be divided into two categories: the fixed subarray and the dynamic subarray. In the fixed subarray architecture, each RF chain is connected with a fixed antenna subset; meanwhile, each antenna is connected to a single RF chain [13], [14]. In dynamic subarray scenario, antenna elements are adaptively partitioned into several subsets based on the long term channel information [15], [16].

The inherent directional characters of mmWave frequencies are beneficial to serve tens of users simultaneously. Thus multi-user hybrid precoding is an important approach to significantly enhance the spectral efficiency and system capacity for mmWave massive MIMO systems. One typical multi-user hybrid precoding scheme designed the analog precoders to harvest the large array gain and the digital Zero forcing (ZF), Minimum mean-squared error (MMSE) or Block diagonalization (BD) processing based on the analog effective channel in [17]-[19]. Another typical multi-user hybrid precoding scheme was designed as a solution of non-orthogonal angle division multiple access based on the angle information extracted from the channel estimation in [20], [21]. Unfortunately, most prior works on multi-user hybrid precoding only considered the fully-connected architecture [17]-[21].

The dynamic subarray achieves a compromise between sum rate and hardware complexity for millimeter wave massive MIMO systems [15]. However, multi-user hybrid precoding in the dynamic subarray architecture is intractable to solve, where the antenna partitioning would result in user unfairness and the multi-user interference (MUI). Limited work has been done for multi-user hybrid precoding in the dynamic subarray architecture. In [22], the antenna partitioning and analog precoder were achieved by the exhaustive search to maximize the analog effective channel gain. Base on the analog precoded channel with low dimension, the digital precoding was utilized to suppress the MUI exploiting ZF criterion. Furthermore, a multi-user analog precoding scheme was proposed in [23]. $N$ antenna elements with the largest amplitude were selected based on the channel of the first user in Multi-user MIMO (MU-MIMO) system. Then, the phase of analog precoder is computed as the quantized phase of the corresponding column vector. The above two sub-steps were carried out iteratively until all MU-MIMO users were completed.

Nevertheless, the aforementioned multi-user hybrid precoding may have the following shortcomings. Firstly, the exhaustive search used in [22] introduces an extremely high complexity, since the effective channel gains for all users with all analog codewords have to be calculated. Then, the solution in [23] is hard to achieve optimal performance as the number of antennas is same connected with each RF chain. Finally, the users order in the procedure of the antenna selection leads to severe unfairness since the first user is able to choose the whole antenna elements and the other users can only choose the remaining elements.

In this paper, a novel multi-user hybrid precoding solution is designed for dynamic subarrays architecture in mmWave Massive MIMO systems. The contributions are summarized as follows:
\begin{itemize}
  \item We propose a multi-user hybrid precoding solution for they dynamic subarray architecture. Firstly, each user selects the best beam to maximize single-user effective channel gains and feedbacks the index to the base station (BS). Then, BS takes it as the initial analog precoder of each user. Secondly, the multi-user set with maximal sum rate is selected. Subsequently, the antennas are partitioned to RF chains based on the maximal signal-to-interference-plus-noise ratio (SINR) increment criterion. Finally, the hybrid precoding scheme is optimized for the dynamic subarray architecture.
  \item We develop an antenna partitioning algorithm for the dynamic subarray. For the selected multi-users, each antenna element is dynamically allocated to RF chain according to the maximal SINR increment. The proposed antenna partitioning algorithm guarantees the user fairness since each antenna element is allocated to acquire the maximal SINR increment of all selected users. Moreover, it can greatly reduce the size of the search space and the calculation complexity because the SINR is calculated on the analog effective channels of the selected multi-users.
\end{itemize}

Simulation results show that the sum rate of the proposed multi-user hybrid precoding solution achieves significant gain compared to the fixed subarray and approaches that of the exhaustive search of the dynamic subarray. The results also confirm the energy efficiency of the proposed solution outperforms the fully-connected architecture because it greatly reduces the number of PSs and eliminates the need for RF adders. Finally, based on the complexity analysis, the computation amount of the proposed antenna partitioning algorithm can be significantly reduced to $N_{\rm{RF}}\times N_{\rm{TX}}$, compared to that of exhaustive search solution in [22], i.e. $\frac{1}{\left ( N_{\rm{RF}} \right )!}\sum_{k=0}^{N_{\rm{RF}}}\left ( -1 \right )^{N_{\rm{RF}}-k}\binom{N_{\rm{RF}}}{k}k^{^{N_{\rm{TX}}}}$.

The remaining parts of this paper are structured as following. Section II provides the description of the system model and channel model. Section III proposes the problem description and section IV illustrates specifics of the proposed solution. The simulation results and the complexity analysis are discussed in section V. Lastly, concluding remarks are presented in section VI.

Notation: Bold uppercase $\mathbf{A}$ is a matrix, $\mathbf{a}$ is a vector, $\mathit{a}$ is a scalar. Moreover, $\mathbf{A}^{H}$, $\mathbf{A}^{-1}$ and $\mathbf{A}^{T}$ are the hermitian operation (conjugate transpose), the inverse operation, and the transpose operation of matrix $\mathbf{A}$, respectively. $\mathbf{I}_{N}$ is the $\mathit{N}$ dimensional identity matrix. $\left \| \mathbf{A} \right \|_{F}$ is Frobenius norm of matrix $\mathbf{A}$. $\mathcal{CN}\left ( \mathbf{m,R} \right )$ is a complex gaussian random vector with mean $\mathbf{m}$ and covariance $\mathbf{R}$. $\mathcal{A}$ is a set.

\section{System Model And Channel Model}

\subsection{System Model}
Consider a downlink multi-user hybrid precoding mmWave system with the conventional fully-connected architecture as shown in Fig. 1, in which BS simultaneously communicates with $K$ mobile users. $N_{\rm{TX}}$ antennas and $N_{\rm{RF}}$ RF chains are equipped in BS such that $N_{\rm{RF}}\leq N_{\rm{TX}}$. Each mobile user configures $N_{\rm{RX}}$ antenna.
\begin{figure}
  \centering
  \includegraphics[width=3.4 in]{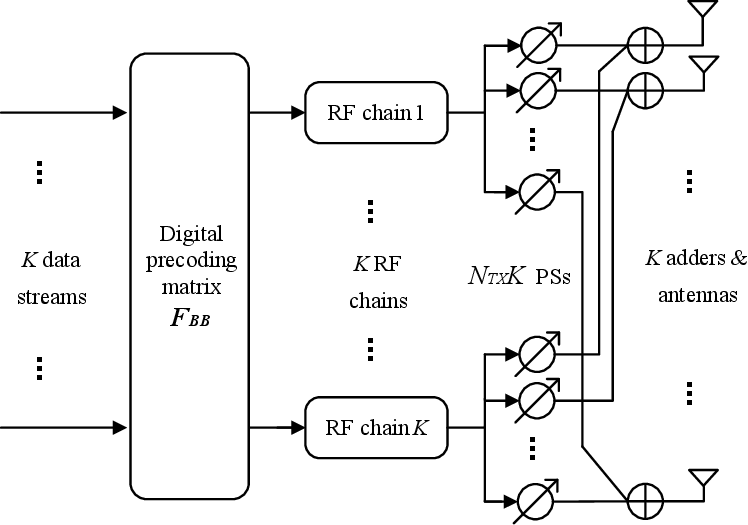}\\
  \caption{Fully-connected architecture in the multi-user mmWave Massive MIMO system.}
\end{figure}

The transmitter adopts a $N_{\rm{RF}}\times N_{s}$ digital precoding weight ${\bf{F}}_{\textrm{BB}}^{\mathit{k}}$, followed by a $N_{\rm{TX}}\times N_{\rm{RF}}$ analog precoding weight, ${{\bf{F}}_{\textrm{RF}}} = \left[ {{\bf{F}}_{\textrm{RF}}^1,{\bf{F}}_{\textrm{RF}}^2, \cdots ,{\bf{F}}_{\textrm{RF}}^K} \right]$, where $\mathbf{F}_{\textrm{RF}}^{\mathit{k}}$ is the analog precoding vector for the $k$th RF chain. $\mathbf{H}_{k}$ denotes the $N_{\rm{RX}}\times N_{\rm{TX}}$ channel matrix from BS to the $k$th mobile user, such that $\mathbb{E}\left[ {\left\| {{{\bf{H}}_k}} \right\|_{\rm{F}}^2} \right] = {N_{{\rm{TX}}}}{N_{{\rm{RX}}}}$. The received signal of mobile user $k$ can be written as

\begin{align}
{{\bf{y}}_k} = \sqrt \rho {{\bf{W}}_k} {{\bf{H}}_k}{\bf{F}}_{\textrm{RF}}{\bf{F}}_{\textrm{BB}}^k{\bf{s}}_k + \sum\limits_{i \ne k}^K {{{\bf{W}}_k}{{\bf{H}}_k}{\bf{F}}_{\textrm{RF}}{\bf{F}}_{\textrm{BB}}^i{\bf{s}}_i}  + {{\bf{W}}_k}{{\bf{n}}_k},
\end{align}where ${\bf{s}}_k$ is the ${N_s} \times 1$ transmitted signal at the $k$th user with $\mathbb{E}\left[ {\bf{s}}_k{\bf{s}}_k^H \right] = {{\bf{I}}_s}$, ${{\bf{n}}_k}\sim\mathcal{CN}\left( {0,\sigma _N^2{\bf{I}}} \right)$ is an additive Gaussian white noise vector with independent and identically distribution $(i.i.d)$ and $\rho $ is defined as the average received power. ${{\bf{W}}_k}$ is the $N_{s}\times N_{\rm{RX}}$ digital matrix at the receiver.

In further, ${{\bf{W}}_k}{{\bf{H}}_k}{\bf{F}}_{\textrm{RF}}{\bf{F}}_{\textrm{BB}}^k{\bf{s}}_k$ represents the desired signal and $\sum\limits_{i \ne k}^K {{{\bf{W}}_k}{{\bf{H}}_k}{\bf{F}}_{\textrm{RF}}{\bf{F}}_{\textrm{BB}}^i{\bf{s}}_i}$ is the multi-user interference (MUI) for the $k$th user, respectively.

The PSs are adopted by the analog precoder, thus the entries of ${\bf{F}}_{{\textrm{RF}}}$ possess constant modulus and are normalized to satisfy ${\left| {{{\left[ {{{\bf{F}}_{\textrm{RF}}}} \right]}_{j,k}}} \right|^2} = 1$, $\left( {j = 1, \ldots ,{N_{\rm{TX}}}} \right)$. The analog precoding codebook $\mathcal{F}$ with cardinality $\left| {\mathcal{F}} \right| = {N_Q}$ is shared by BS and the user equipment. On the condition of beam steering codebooks, $\mathcal{F}$ consists of the vectors ${{\bf{b}}_q} = \left[ {1,{e^{j\frac{{2\pi }}{\lambda }dsin\left( {\frac{{2\pi q}}{{{N_Q}}}} \right)}}, \cdots ,{e^{j\left( {{N_{\rm{TX}}} - 1} \right)\frac{{2\pi }}{\lambda }dsin\left( {\frac{{2\pi q}}{{{N_Q}}}} \right)}}} \right]$, where the variable $q$ taking the values $0,1,2$ and ${N_Q} - 1$. The total power of transmitter is constrained to $\left\| {{\bf{F}}_{\textrm{RF}}^{\mathit{k}}{\bf{F}}_{\textrm{BB}}^{\mathit{k}}} \right\|_F^2 = {N_s}$ by normalizing ${\bf{F}}_{\textrm{BB}}^{\mathit{k}}$.

The multi-user hybrid precoding in dynamic subarray architecture is shown in Fig. 2, where each RF chain dynamically connects to a subset of large-scale antenna elements and each antenna is only connected to one RF chain as defined in [14], [15]. Assume BS transmits data to every mobile user via only one stream considering the inherent directionality of mmWave. Therefore, the number of MU-MIMO users is equal to the number of RF chains, i.e. $K= N_{\rm{RF}}$. From $K= N_{\rm{RF}}$, the $k$th user is mapped to the $k$th RF chain with the analog precoding vector.

Correspondingly, ${\mathcal{S}_k}$ denotes the antenna subarray connected to the $k$th RF chain. The subarray ${\mathcal{S}_k}$ is comprised by $N_{\rm{k}}$ antenna elements such that $1 \le {N_{\rm{k}}} < {N_{\rm{TX}}}$ and ${N_{\rm{TX}}} = \sum\limits_{k = 1}^K {{N_{\rm{k}}}} $.
\begin{figure}
  \centering
  \includegraphics[width=3.4 in]{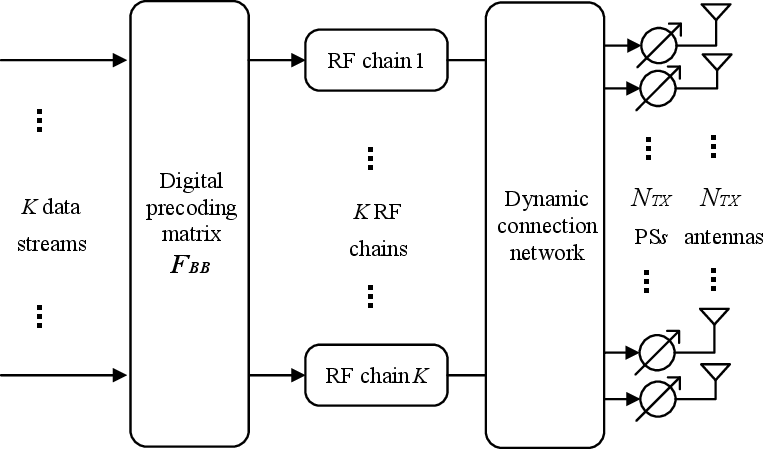}\\
  \caption{Dynamic subarray architecture in the multi-user mmWave Massive MIMO system.}
\end{figure}

According to the dynamic subarray architecture as aforementioned, the received signal of mobile user $k$ can be rewritten as

\begin{align}
\begin{array}{l}
{{\bf{y}}_{{\mathcal{S}_k}}} = \sqrt \rho  {{{\bf{W}}_{\mathcal{S}_k}}{\bf{H}}_{{\mathcal{S}_k}}}{\bf{F}}_{\textrm{RF}}^{{\mathcal{S}_k}}{\bf{F}}_{\textrm{BB}}^{{\mathcal{S}_k}}{{\bf{s}}_k} + \sum\limits_{i \ne k}^K {{{\bf{W}}_{\mathcal{S}_k}}{{\bf{H}}_{{\mathcal{S}_k}}}{\bf{F}}_{\textrm{RF}}^{{\mathcal{S}_i}}{\bf{F}}_{\textrm{BB}}^{{\mathcal{S}_i}}{{\bf{s}}_i}} \\
+ {{\bf{W}}_{\mathcal{S}_k}}{{\bf{n}}_k},
\end{array}
\end{align}where ${{\bf{W}}_{\mathcal{S}_k}}$ is the $N_{s}\times N_{\rm{RX}}$ digital matrix at the receiver. ${\bf{H}}_{\mathcal{S}_k}$ is the ${N_{\rm{RX}}} \times {N_k}$ channel matrix from the subarray ${\mathcal{S}_k}$ at the transmitter to the $k$th mobile user, ${\bf{F}}_{\textrm{RF}}^{\mathcal{S}_k}$ is the ${N_k} \times {N_{\rm{RF}}}$ analog precoding vector connecting the $k$th RF chain and the subarray ${\mathcal{S}_k}$. ${\bf{F}}_{\textrm{BB}}^{\mathcal{S}_k}$ is the ${N_{\rm{RF}}} \times {N_s}$ digital precoding vector of the $k$th user.

\subsection{Channel Model}
The severe pathloss in mmWave frequency leads to limited scattering [2], [4]. The geometric channel model is adopted with ${L_k}$ scatterers for the $k$th user. It is assumed that each scatterer contributes a single propagation path between the BS and user [5], [24], [25]. The channel of each mobile user $k$ can be expressed as

\begin{align}
{{\bf{H}}_k} = \sqrt {\frac{{{N_{\rm{TX}}}{N_{\rm{RX}}}}}{{{L_k}}}} \sum\limits_{l = 1}^{{L_k}} {{\alpha _{k,l}}} {\bf{a}}_{\textrm{MS}}^H\left( {{\theta _{k,l}}} \right){{\bf{a}}_{\textrm{BS}}}\left( {{\phi _{k,l}}} \right),
\end{align}where ${\alpha _{k.l}}$ is the complex gain of the $l$th path between the BS and the $k$th user, ${L_k}$ is the number of scatterers. The variable ${\theta _{k,l}} \in \left[ {0,2\pi } \right]$ represents the azimuth angle of arrival (AOA) of the $l$th path. ${\phi _{k,l}} \in \left[ {0,2\pi } \right]$ represents the azimuth angle of departure (AoD) of the $l$th path. Consequently, ${{\bf{a}}_{\textrm{MS}}}\left( {{\theta _{k,l}}} \right)$ and ${{\bf{a}}_{\textrm{BS}}}\left( {{\phi _{k,l}}} \right)$ represent the array response vectors of the $k$th user and the BS respectively. The uniform linear arrays (ULAs) are used, and the array response vector ${{\bf{a}}_{\textrm{BS}}}\left( {{\phi _{k,l}}} \right)$ of the BS is

\begin{align}
{{\bf{a}}_{\textrm{BS}}}\left( {{\phi _{k,l}}} \right) = {\left[ {\begin{array}{*{20}{c}}
1\\
{{e^{j\frac{{2\pi }}{\lambda }dsin\left( {{\phi _{k,l}}} \right)}}}\\
 \vdots \\
{{e^{j\left( {{N_{\rm{TX}}} - 1} \right)\frac{{2\pi }}{\lambda }dsin\left( {{\phi _{k,l}}} \right)}}}
\end{array}} \right]^T},
\end{align} where $d$ is the spacing distance between two adjacent antenna elements, and $\lambda$ is the wavelength of transmitting signals. ${{\bf{a}}_{\textrm{MS}}}\left( {{\theta _{k,l}}} \right)$ can be formulated in a similar fashion.

For dynamic subarray architecture, the channel of mobile user $k$ can be represented as

\begin{align}
{{\bf{H}}_{{\mathcal{S}_k}}} = \sqrt {\frac{{{N_k}{N_{\rm{RX}}}}}{{{L_k}}}} \sum\limits_{l = 1}^{{L_k}} {{\alpha _{k,l}}} {\bf{a}}_{\textrm{MS}}^H\left( {{\theta _{k,l}}} \right){\bf{a}}_{\textrm{BS}}^{{\mathcal{S}_k}}\left( {{\phi _{k,l}}} \right),
\end{align}where array response vector ${\bf{a}}_{\textrm{BS}}^{{\mathcal{S}_k}}\left( {{\phi _{k,l}}} \right)$ corresponding to the antenna set ${\mathcal{S}_k}$ can be given by

\begin{align}
{\bf{a}}_{\rm{BS}}^{{\mathcal{S}_k}}\left( {{\phi _{k,l}}} \right) = {\left[ {\begin{array}{*{20}{c}}
{{e^{j\left( {\mathcal{S}_k^1 - 1} \right)\frac{{2\pi }}{\lambda }dsin\left( {{\phi _{k,l}}} \right)}}}\\
 \vdots \\
{{e^{j\left( {\mathcal{S}_k^i - 1} \right)\frac{{2\pi }}{\lambda }dsin\left( {{\phi _{k,l}}} \right)}}}\\
 \vdots \\
{{e^{j\left( {\mathcal{S}_k^{{N_k}} - 1} \right)\frac{{2\pi }}{\lambda }dsin\left( {{\phi _{k,l}}} \right)}}}
\end{array}} \right]^T},
\end{align} where $\mathcal{S}_k^i$ is the $i$th antenna index of ${\mathcal{S}_k}$ expressed by $\left\{ {\mathcal{S}_k^1, \ldots ,\mathcal{S}_k^i, \ldots ,\mathcal{S}_k^{{N_k}}} \right\}$.

\section{Problem Formulation}

Given the system model for the dynamic subarray architecture in eq. (2), the achievable rate of the $k$th user corresponding to the antenna set ${\mathcal{S}_k}$ is written as

\begin{align}
{R_k} = {\log _2}\left( {1 + \frac{{\left\| {{{\bf{W}}_{{\mathcal{S}}_k}}}{{{\bf{H}}_{{{\mathcal{S}}_k}}}{\bf{F}}_{\textrm{RF}}^{{{\mathcal{S}}_k}}{\bf{F}}_{\textrm{BB}}^{{{\cal S}_k}}} \right\|_F^2}}{{\sigma _n^2 + \sum\nolimits_{i \ne k}^K {\left\| {{{\bf{W}}_{{\mathcal{S}}_k}}}{{{\bf{H}}_{{{\mathcal{S}}_k}}}{\bf{F}}_{\textrm{RF}}^{{{\mathcal{S}}_i}}{\bf{F}}_{\textrm{BB}}^{{{\mathcal{S}}_i}}} \right\|_F^2} }}} \right).
\end{align}

The main objective of this paper is to design the multi-user hybrid precoding weights for the dynamic subarray architecture $\left\{ {{\bf{F}}_{\textrm{RF}}^*,{\bf{F}}_{\textrm{BB}}^*,{\mathcal{S}}_k^*} \right\}$ at the transmitter. In this paper, ${{\bf{W}}_k} = {\bf{U}}_k^H$, where ${\bf{U}}_k^H$ is obtained by the Singular value decomposition (SVD) of the channel matrix ${{\bf{H}}_{{\mathcal{S}_k}}}$, which ${{\bf{H}}_{{\mathcal{S}_k}}} = {{\bf{U}}_k}{{\bf{\Lambda }}_k}{\bf{V}}_k^H$. For the sake of simplicity, let ${{\bf{\tilde H}}_{{\mathcal{S}_k}}}$ represent ${{\bf{W}}_k}{{\bf{H}}_{{\mathcal{S}_k}}}$ in the following. Then the objective function of hybrid precoding for dynamic subarray is described as

\begin{align}
\begin{array}{l}
\left\{ {{\bf{F}}_{\textrm{RF}}^*,{\bf{F}}_{\textrm{BB}}^*,\mathcal{S}_k^*} \right\} = \mathop {\arg \max }\limits_{{{\bf{F}}_{\textrm{RF}}},{{\bf{F}}_{\textrm{BB}}},{\mathcal{S}_k}} \sum\limits_{k = 1}^K {{R_k}}, \\
s.t.\left\| {{{\left[ {{\bf{F}}_{\textrm{RF}}^{}} \right]}_{j,k}}} \right\|_F^2 = 1,j = 1, \ldots ,{N_{\rm{TX}}},\\
\begin{array}{*{20}{c}}
{}&{\left\| {{\bf{F}}_{\textrm{RF}}^{}{\bf{F}}_{\textrm{BB}}^{}} \right\|_F^2 = {N_s}},
\end{array}
\end{array}
\end{align}which is a joint optimization problem to the three matrix variables $\left\{ {{\bf{F}}_{\textrm{RF}}^*,{\bf{F}}_{\textrm{BB}}^*,{\mathcal{S}}_k^*} \right\}$. Unfortunately, this problem is found to be intractable to acquire the global optima for joint optimization problems with the similar constrains [26], [27]. Thus the non-convex constraints on the hybrid precoding $\left\{ {{\bf{F}}_{\textrm{RF}}^*,{\bf{F}}_{\textrm{BB}}^*,{\mathcal{S}}_k^*} \right\}$ are impossible to be directly solved. To simplify the hybrid precoding design of the dynamic subarray, the optimization problem is temporarily decoupled and decomposed into three simple maximal sum rate optimization problems for $\left\{ {{\bf{F}}_{\textrm{RF}}^*} \right\}$, $\left\{ {\mathcal{S}_k^*} \right\}$ and $\left\{ {{\bf{F}}_{\textrm{BB}}^*} \right\}$, respectively.

The detail of the proposed solution will be explained in section IV.

\section{The Proposed Method}
In this section, a general framework for multi-user hybrid precoding in dynamic subarray system is designed. The joint optimization problem $\left\{ {{\bf{F}}_{\textrm{RF}}^*,{\bf{F}}_{\textrm{BB}}^*,{\mathcal{S}}_k^*} \right\}$ is decomposed into multiple sub-problems which includes the analog precoding initialization, the multi-user selection, the dynamic subarray partitioning, and the optimization of hybrid precoding for the dynamic subarray architecture. The main idea of the proposed solution can be summarized as follows:

\begin{enumerate}
  \item Each user searches the best beam from the codebook which can obtain the maximal single-user analog effective channel gain and feedbacks the index of the best beam to BS. Accordingly, BS takes the best beam as the initial analog precoding vector of each user.
  \item Then, exploiting the initial analog effective channel, BS selects the multi-user set from $N$ candidate users to maximize the sum rate.
  \item For the selected multi-user set, the antenna partitioning algorithm is designed to maximize SINR increment of all selected multi-users.
  \item The analog precoding vector is solved for the dynamic subarray architecture, and the ZF linear precoding is adopted as digital precoding to eliminate MUI, respectively.
\end{enumerate}
\subsection{Analog Precoding}
MmWave channel possesses inherent directionality characteristics, thus the best beams of the candidate users depend on their own scattering paths. For the single user hybrid precoding, the common approach of the analog precoding is searching the strongest beam in the whole codebook [4], [6]. For multi-user hybrid precoding, the space division multiplexing is an important approach to mitigate MUI in mmWave system [20], [21]. Therefore, the best beam of each active user is firstly selected based on its own channel and utilized as the important information for multi-user hybrid precoding. That is, BS adopts the best beam of each candidate user as its initial analog precoding.

The process to solve the initial analog precoding is constructed specifically as:

Before the downlink transmission, BS broadcasts the reference signals sequentially precoded by the codeword of the codebook. Then the user measures the power of reference signals and selects the codeword of the strongest receiving reference signals as the best beam. The index of the best beam is sent back to the BS. At the last step, BS takes the corresponding codeword as the initial analog precoding vector of this user.

In further, the analog precoding vector of the $n$th user can be selected from the analog precoding codebook $\mathcal{F}$ according to the following criterion:

\begin{align}
{\bf{f}}_n^o = \mathop {\arg \max }\limits_{{{\bf{b}}_q} \in \mathcal{F}} \left\| {{{\bf{\tilde H}}_n}{{\bf{b}}_q}} \right\|_F^2,
\end{align}where ${{\bf{b}}_q}$ is the $q$th codeword from the analog precoding codebook $\mathcal{F}$. ${{\bf{\tilde H}}_n}$ represents the effective channel ${{\bf{W}}_n}{{\bf{H}}_n}$ of the $n$th user. ${\bf{f}}_n^o \in {{\bf{C}}^{{N_{{\rm{TX}}}} \times 1}}$ indicates the initial analog precoding vector to the $n$th user ($n = 1,2, \ldots ,N$) and $N$ is the number of candidate users.

\subsection{MU-MIMO User Selection}
When the BS transmits signals to multiple users in the same time slot, MUI severely degrades the system performance. The aim of multi-user selection is to select a group of MU-MIMO users with minimal inter-user interference and maximal objective channel gains. Usually maximizing the sum rate is the criterion of user selection, the SINR for each user should be estimated by scheduler to simplify the calculation [30]. The basic principle is that only if user has maximal SINR value, it is added to the selected user set as described in [28], [29].

Exploiting the initial analog precoding vectors, the SINR of the $n$th user is written as

\begin{align}
{\rm{SINR}_n} = \frac{{\left\| {{{\bf{\tilde H}}_n}{\bf{f}}_n^o} \right\|_F^2}}{{{\sigma _n^2} + \sum\nolimits_{i \ne n}^N {\left\| {{{\bf{\tilde H}}_n}{\bf{f}}_i^o} \right\|} _F^2}}.
\end{align}

The process of the multi-user selection is constructed as:

Let us define a set of all candidate users $\mathcal{T} = \left[ {1,2, \ldots ,N} \right]$ and an empty set $\mathcal{U}$ which is updated as the selected multi-user set. The algorithm selects the first user $\mathcal{U}\left( 1 \right)$ with maximum channel gain $\left\| {{{\bf{\tilde H}}_{\mathcal{T}\left( n \right)}}} \right\|_F^2$ from the set $\mathcal{T}$. Then the set $\mathcal{T}$ and the set $\mathcal{U}$ are updated as following

\begin{align}
{\mathcal{T} \leftarrow \mathcal{T}\backslash \mathcal{U}\left( 1 \right),\mathcal{U} \leftarrow \mathcal{U} \cup \mathcal{U}\left( 1 \right)}.
\end{align}

According to the maximal SINR criterion, the new user is included to the selected multi-user set successively. The algorithm selects the remaining users from the set $\mathcal{T}$. And the $k$th user which will be added into the set $\mathcal{U}$ can be represented by following expression.

\begin{align}
\begin{array}{l}
\mathcal{U}\left( k \right) = \mathop {\arg \max }\limits_{\mathcal{T}\left( n \right)} {\rm{SINR}}\left[ {\mathcal{T}\left( n \right)} \right]\\
 = \mathop {\arg \max }\limits_{\mathcal{T}\left( n \right)} \frac{{\left\| {{{\bf{\tilde H}}_{\mathcal{T}\left( n \right)}}{\bf{f}}_{\mathcal{T}\left( n \right)}^o} \right\|_F^2}}{{\sigma _n^2 + \sum\nolimits_{i = 1}^{k - 1} {\left\| {{{\bf{\tilde H}}_{\mathcal{T}\left( n \right)}}{\bf{f}}_{\mathcal{U}\left( i \right)}^o} \right\|} _F^2}},\\
k = 2, \ldots ,{N_{{\rm{RF}}}},
\end{array}
\end{align}where ${\bf{f}}_{\mathcal{T}\left( n \right)}^o$ is the initial analog precoding vector of the $n$th user in the set $\mathcal{T}$, ${\bf{f}}_{\mathcal{U}\left( i \right)}^o$ is the initial analog precoding vector of the $i$th user from set $\mathcal{U}$, and ${\bf{\tilde H}}_{\mathcal{T}\left( n \right)}$ is the effective channel of the $n$th user from set $\mathcal{T}$. Then set $\mathcal{T}$ and set $\mathcal{U}$ can be updated by

\begin{align}
{\mathcal{T} \leftarrow \mathcal{T}\backslash \mathcal{T}\left( n \right),\mathcal{U} \leftarrow \mathcal{U} \cup \mathcal{T}\left( n \right)}.
\end{align}

In each loop, the user with maximal SINR value is added to $\mathcal{U}$ and gets removed from $\mathcal{T}$. The process continues until the set $\mathcal{U}$ contains $K$ users.

\subsection{Dynamic sub-array partitioning algorithm}
In order to address the trade-off between the achievable spectral efficiency and hardware complexity, the dynamic subarrays dynamically partition antenna elements to RF chain based on the long-term channel information [15], [16]. Different from the single-user case in [15], [16], we design the antenna partitioning algorithm for multi-user dynamic subarray hybrid architecture as defined in subsection II. A.

For the antenna subarray $\mathcal{S}_k$, the SINR of the $k$th user is represented by following expression,

\begin{align}
{\rm{SINR}}\left[ {{\mathcal{S}_k}} \right] = \frac{{\left\| {{{\bf{\tilde H}}_{{\mathcal{S}_k}}}{\bf{F}}_{\textrm{RF}}^{{\mathcal{S}_k}}{\bf{F}}_{\textrm{BB}}^{{\mathcal{S}_k}}} \right\|_F^2}}{{\sigma _n^2 + \sum\nolimits_{i \ne k}^K {\left\| {{{\bf{\tilde H}}_{{\mathcal{S}_k}}}{\bf{F}}_{\textrm{RF}}^{{\mathcal{S}_i}}{\bf{F}}_{\textrm{BB}}^{{\mathcal{S}_i}}} \right\|_F^2} }}.
\end{align}

To maximize the sum rate of MU-MIMO user, the dynamic subarray is partitioned according to the maximal SINR of the selected MU-MIMO users. The exhaustive search from all probable cases of the three unknown matrix variables $\left\{ {{\bf{F}}_{\textrm{RF}}^*,{\bf{F}}_{\textrm{BB}}^*,\mathcal{S}_k^*} \right\}$ could be the direct solution for the dynamic subarray but lead to high computational complexity. To address this issue, the analog precoding vectors are exploited as the important information to partition antenna elements. It enables each selected user to fully take advantage of large-scale array gains generated by the directional transmission of mmWave Massive MIMO systems. The optimal antenna subarray ${\mathcal{S}_k^*}$ for the $k$th user is represented by following expression

\begin{align}
{\mathcal{S}_k^* = \mathop {\arg \max }\limits_{{\mathcal{S}_k}} \frac{{\left\| {{{\bf{\tilde H}}_{{\mathcal{S}_k}}}{\bf{f}}_{\mathcal{S}_k}^o} \right\|_F^2}}{{\sigma _n^2 + \sum\nolimits_{i \ne k}^K {\left\| {{{\bf{\tilde H}}_{{\mathcal{S}_k}}}{\bf{f}}_{\mathcal{S}_i}^o} \right\|_F^2} }}},
\end{align} where ${\bf{f}}_{{\mathcal{S}_k}}^o \in {C^{{N_k} \times 1}}$ is obtained by selecting the values of beam ${\bf{f}}_k^o$ according to the dynamic subarray effective channel ${{\bf{\tilde H}}_{{\mathcal{S}_k}}}$. ${\bf{f}}_{{\mathcal{S}_k}}^o$ can be written as

\begin{align}
{\bf{f}}_{{\mathcal{S}_k}}^o = {\left[ {\begin{array}{*{20}{c}}
{{e^{j\left( {\mathcal{S}_k^1 - 1} \right)\frac{{2\pi }}{\lambda }dsin\left( {\frac{{2\pi {q_k}}}{{{N_Q}}}} \right)}}}\\
 \vdots \\
{{e^{j\left( {\mathcal{S}_k^{{N_k}} - 1} \right)\frac{{2\pi }}{\lambda }dsin\left( {\frac{{2\pi {q_k}}}{{{N_Q}}}} \right)}}}
\end{array}} \right]^T},
\end{align} where $\left\{ {\mathcal{S}_k^1, \ldots ,\mathcal{S}_k^{{N_k}}} \right\}$ is the antenna index for the dynamic subarray ${\mathcal{S}_k}$ of the $k$th user, ${q_k}$ is the label of the best beam in the codebook ${\mathcal{F}}$ which was selected for the $k$th user in Subsection IV. A.

However, the maximal SINR criterion results in the severe user unfairness on account that the SINR value is higher for the user being partitioned more antennas. On the other hand, the first antenna generates more SINR increment than other antenna, and so on [3]. Considering the user fairness and the objective function of the maximal sum rate criterion, the dynamic subarray allocates each antenna element to RF chain according to the maximal SINR increment. The SINR increment ${\nabla _k}$ of antenna subarray ${\mathcal{S}_k}$ can be defined as

\begin{align}
{\nabla _k} = {\rm{SINR}}[{\mathcal{S}_k} \cup j] - {\rm{SINR}}[{\mathcal{S}_k}],
\end{align} where $[{\mathcal{S}_k} \cup j]$ represents that the antenna $j$ is added in subarray ${\mathcal{S}_k}$. Thus, the optimal subarray $\mathcal{S}_k^*$ can be rewritten as

\begin{align}
\mathcal{S}_k^* = \mathop {\arg \max }\limits_{{\mathcal{S}_k}} {\nabla _k}.
\end{align}

The algorithm is circularly performed as the following process:

At the initial stage, the dynamic subarray of each user is an empty set and the candidate antenna set contains all antenna elements. Then, the algorithm updates SINR and SINR increment values of each user respectively when an antenna is added into the dynamic subarray. At last, the algorithm finds the subarray $\mathcal{S}_k^*$ with the maximal SINR increment value and assigns this antenna element to the optimal subarray $\mathcal{S}_k^*$. Note that only one antenna is assigned and other antennas remain unchanged at each antenna selection stage. The above process is performed iteratively until all antennas are assigned.

In this paper, the antenna number $N_k$ of subarray $\mathcal{S}_k$ is adaptive to the channel state in order that multi-users can obtain more array gains. First, the calculation complexity can greatly be reduced since the SINR is calculated on the analog effective channel with low dimension. Further, the selection of each antenna element guarantees user fairness because the SINR increment is maximal for all MU-MIMO users. At last, the number of iterations is significantly decreased as it is equal to the amount of multi-users instead of all candidate users.

The process of the proposed algorithm is described in Algorithm 1.

\noindent
\begin{tabular}{lcl}
\\  \toprule
$\bf{Algorithm\ 1}$: Dynamic sub-array partitioning  \\ \midrule
$\bf{Input}$: ${N_{{\rm{TX}}}}, {K}, {\mathcal{S}_0} = \left\{ {1, \ldots ,{N_{{\rm{TX}}}}} \right\}, {{\mathcal{S}_1}, \ldots ,{\mathcal{S}_K} = \phi }$\\
for $j = 1:{N_{{\rm{TX}}}}$\\
~~~for $k = 1:K$\\
~~~~~~~${\rm{SINR}}\left[ {\mathcal{S}_k} \right] = \frac{{\left\| {{{{\bf{\tilde H}}}_{{\mathcal{S}_k}}}{{\bf{f}}}_{{{\mathcal{S}}_k}}^o} \right\|_F^2}}{{\sigma _n^2 + \sum\nolimits_{i \ne k}^K {\left\| {{{\bf{\tilde H}}_{{{\mathcal{S}}_k}}}{\bf{f}}_{{\mathcal{S}_i}}^o} \right\|_F^2} }}$\\
~~~~~~~${\rm{SINR}}\left[ {{\mathcal{S}_k} \cup j} \right] = \frac{{\left\| {{{\bf{\tilde H}}_{{\mathcal{S}_k} \cup j}}{\bf{f}}_{{\mathcal{S}_k} \cup j}^o} \right\|_F^2}}{{\sigma _n^2 + \sum\nolimits_{i \ne k}^K {\left\| {{{\bf{\tilde H}}_{{\mathcal{S}_k} \cup j}}{\bf{f}}_{{\mathcal{S}_i}}^o} \right\|_F^2} }}$\\
~~~~~~~${\nabla _k} = {\rm{SINR}}[{\mathcal{S}_k} \cup j] - {\rm{SINR}}[{\mathcal{S}_k}]$\\
~~~end\\
~~~$\mathcal{S}_k^* = \mathop {\arg \max }\limits_{{\mathcal{S}_k}} {\nabla _k}$\\
~~~$\mathcal{S}_k^* \leftarrow \mathcal{S}_k^* \cup j,{\mathcal{S}_0} \leftarrow {\mathcal{S}_0}\backslash j$\\
end\\
$\bf{Output}$:${\mathcal{S}_1}, \ldots ,{\mathcal{S}_{{N_{{\rm{RF}}}}}}$
\\\bottomrule
\end{tabular}

\subsection{Hybrid Precoding}
After the partitioning of dynamic subarrays, the optimization problem in eq. (8) actually becomes similar to the conventional hybrid precoding problem.

The only difference is the dynamic subarray architecture. Both the beam shape and width of each dynamic sub-array are changed as the antenna elements which are different from the full-connected architecture. Thus the initial analog precoding should be updated by the dynamic subarray. For the dynamic subarray with the partitioned antenna elements, the vector ${\bf{b}}_q^{{\mathcal{S}_k}}$ in codebook ${\mathcal{F}^{{\mathcal{S}_k}}}$ should be rewritten:

\begin{align}
{\bf{b}}_q^{{\mathcal{S}_k}} = {\left[ {\begin{array}{*{20}{c}}
{{e^{j\left( {\mathcal{S}_k^1 - 1} \right)\frac{{2\pi }}{\lambda }dsin\left( {\frac{{2\pi q}}{{{N_Q}}}} \right)}}}\\
 \vdots \\
{{e^{j\left( {\mathcal{S}_k^{{N_k}} - 1} \right)\frac{{2\pi }}{\lambda }dsin\left( {\frac{{2\pi q}}{{{N_Q}}}} \right)}}}
\end{array}} \right]^T},
\end{align} where $\left\{ {\mathcal{S}_k^1, \ldots ,\mathcal{S}_k^{{N_k}}} \right\}$ is the antenna index for the dynamic subarray $\mathcal{S}_k$ of each user.

Here, the analog precoding vector ${\bf{F}}_{{\textrm{RF}}}^{{\mathcal{S}_k}}$ for the dynamic subarray $\mathcal{S}_k$ is determined to:

\begin{align}
{\bf{F}}_{{\textrm{RF}}}^{{\mathcal{S}_k}} = \mathop {{\rm{argmax}}}\limits_{{\bf{b}}_q^{{\mathcal{S}_k}} \in {\mathcal{F}^{{\mathcal{S}_k}}}} \left\| {{{\bf{\tilde H}}_{{\mathcal{S}_k}}}{\bf{b}}_q^{{\mathcal{S}_k}}} \right\|_F^2,
\end{align} where ${\bf{b}}_q^{{\mathcal{S}_k}}$ is the $q$th codeword of the analog precoding codebook ${\mathcal{F}^{{\mathcal{S}_k}}}$.

Then, the aim of digital precoding is to eliminate the inter-user interference according to the maximal SINR criterion. The digital precoding algorithms are adopted as the classical ZF and MF schemes [3], [4]

\begin{align}
{\bf{F}}_{{\textrm{BB}}}^{{\textrm{ZF}}}\left( k \right) = {\bf{\bar H}}_{{\mathcal{S}_k}}^H{\left( {{{{\bf{\bar H}}}_{{\mathcal{S}_k}}}{\bf{\bar H}}_{{\mathcal{S}_k}}^H} \right)^{ - 1}},
\end{align}

\begin{align}
{\bf{F}}_{{\rm{BB}}}^{{\rm{MF}}}\left( k \right) = \frac{{{\bf{\bar H}}_{{\mathcal{S}_k}}^H}}{{\left\| {{\bf{\bar H}}_{{\mathcal{S}_k}}^H} \right\|_F^2}},
\end{align} where ${{\bf{\bar H}}_{{\mathcal{S}_k}}} = {{\bf{\tilde H}}_{{\mathcal{S}_k}}}{\bf{F}}_{{\rm{RF}}}^{{\mathcal{S}_k}}$ is the analog effective channel of the $k$th user.

\section{Simulation Results}
In this section, the performance of the proposed solution is evaluated by extensive computer simulations. The exhaustive search of dynamic subarray, the fixed subarray and the conventional full-connected architecture are chosen as benchmarks. In order to validate the superiority of the proposed solution, we compare the sum rate and the energy efficiency of three array architectures, respectively. At last, the computational complexity of the proposed solution is investigated.

Specially, two kinds of the fixed subarray cases are adopted as shown in Fig. 3, e.g. the adjacent structure and the interlaced structure where $m = {{{N_{{\rm{TX}}}}} \mathord{\left/
 {\vphantom {{{N_{{\rm{TX}}}}} {{N_{{\rm{RF}}}}}}} \right.
 \kern-\nulldelimiterspace} {{N_{{\rm{RF}}}}}}$. For the exhaustive search algorithm of the dynamic subarray architecture, the optimal subarrays are found by the exhaustive search over all the antenna elements and the analog precoding codewords as described in [22]. Considering the fairness comparison, the full-connected architecture and the fixed subarray architecture adopt the same hybrid precoding method in simulations as proposed in [13].

 \begin{figure}
  \centering
  \includegraphics[width=3.4 in]{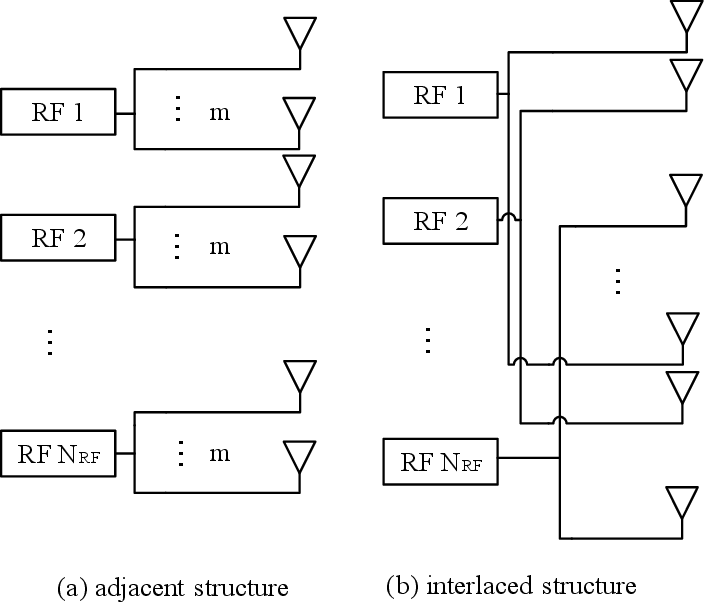}\\
  \caption{Two structures of the fixed subarray using different mapping strategies: each RF chain is connected to adjacent antennas in (a) and
to interlaced antennas in (b).}
\end{figure}

Without loss of generality, the key simulation parameters are the same as those in [15], [17] and are listed in Table I. In the simulations, the geometric channel model with ${L_k}$ scatterers is adopted as described in Subsection II. B. The distributions of the paths delay and the azimuth angles are similar to that in WINNER II SCM channel model [25].

\begin{table}[!htp]
\begin{center}
\caption{Simulation Parameters}
\begin{tabular}{|c|c|}
  \hline
  Number of antennas at BS, $N_{\rm{TX}}$ & 16, 32, 64, 128, 256 \\\hline
  Number of antennas at user, $N_{\rm{RX}}$ & 2 \\\hline
  Number of users, $K$ & $2 \le K \le 8 (K = {N_{\rm{RF}}})$ \\\hline
  Number of scatterers, ${L_k}$ & $4$ \\\hline
  Range of azimuth angle  & uniformly distribution in \\ & $\left[ { - {{180}^ \circ },{{180}^ \circ }} \right]$ \\\hline
  Size of codebook  &   32   \\\hline
  Antenna spacing  & $0.5\lambda $ \\\hline
  Carrier frequency  & $60{\rm{GHz}}$ \\\hline
  Power consumption of RF chain, ${P_{\rm{RF}}}$  &   $250{\rm{mW}}$ \\\hline
  Power consumption of PS, ${P_{\rm{PS}}}$   &   $1{\rm{mW}}$ \\\hline
  Power amplifier efficiency, $\eta $                &   0.38 \\\hline
\end{tabular}
\end{center}
\end{table}

\subsection{Performance Comparisons of Sum Rate}
Firstly, we investigate the sum rate of the proposed solution and the benchmark schemes. BS is configured with 64 antennas (Uniform Linear Array, ULA) and 2 RF chains serving two users simultaneously. As observed from Fig. 4, the sum rate of the proposed solution significantly outperforms two kinds of the fixed subarrays because the antenna elements in the proposed solution are adaptively partitioned to RF chains according to the long-term channel information. And then, the sum rates of the proposed multi-user hybrid precoding scheme approach to that of the exhaustive search in dynamic subarray architecture with lower calculation complexity. At last, the result shows that the performance loss of the proposed solution is negligible to compare with the full-connected architecture. For instance, the proposed solution can obtain about 97.8\% of the sum rate achieved by full-connected hybrid architecture and more than 7\% of the sum rate achieved by the fixed subarray architecture at SNR = 0 dB.

\begin{figure}
  \centering
  \includegraphics[width=3.7 in]{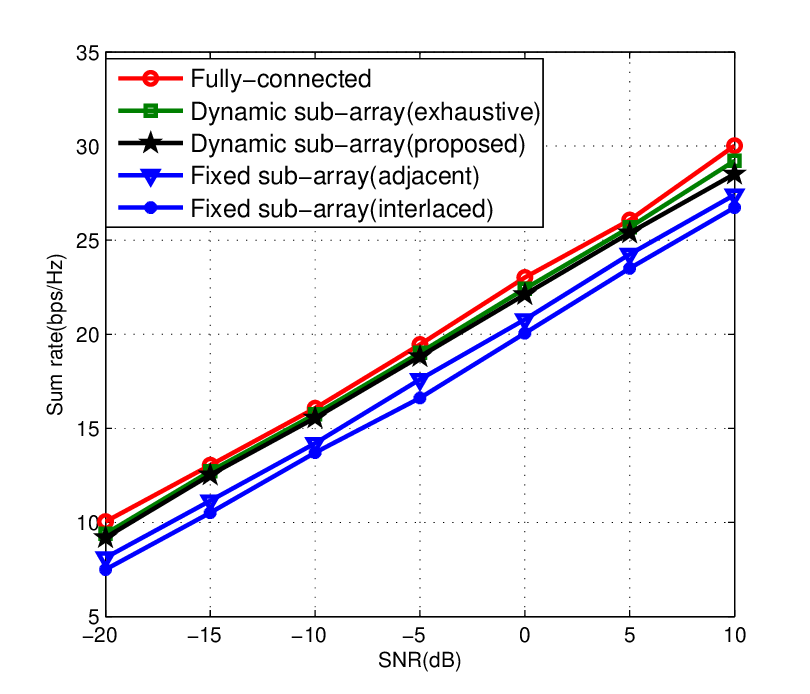}\\
  \caption{Performance comparisons of sum rate for various array architectures when BS equips 64 antennas (ULA) and 2 RF chains.}
\end{figure}

In Fig. 5, we investigate the performance of three array architectures for diverse number of RF chains at the transmitter. The figures show the sum rate versus SNR in the case of 64 and 128 transmit antennas and two users with 2 receive antennas. The number of RF chains at BS is 2, 4, or 8, respectively. The fixed subarray adopts the interlaced architecture in this simulation. Observed from Fig. 5 (a), Fig. 5 (b) and Fig. 5 (c), the dynamic subarray obviously performs better than the fixed subarray in diverse settings because the adaptive antenna partitioning algorithm obtains more array processing gains.

More importantly, the performance gaps of three array architectures are more obvious with 8 RF chains observed from Fig. 5 (c). It validates our analysis that the dynamic subarray would bear the slight performance degradation with lower power consumption and hardware cost compared with the full-connected architecture. For example, when BS is configured with 128 antennas and 8 RF chains, the number of the PSs and RF adders can be reduce to 128 and 0 in the dynamic subarray and the fixed subarray compared with $8 \times 128$ and 128 in the full-connected architecture, respectively.

\textcolor{blue}{\begin{figure}[!htbp]
\centering
\subfigure[Sum rate vs. SNR (64, 128 antennas and 2 RF chains).]{
\label{fig:subfig:a}
\includegraphics[width=3.4 in]{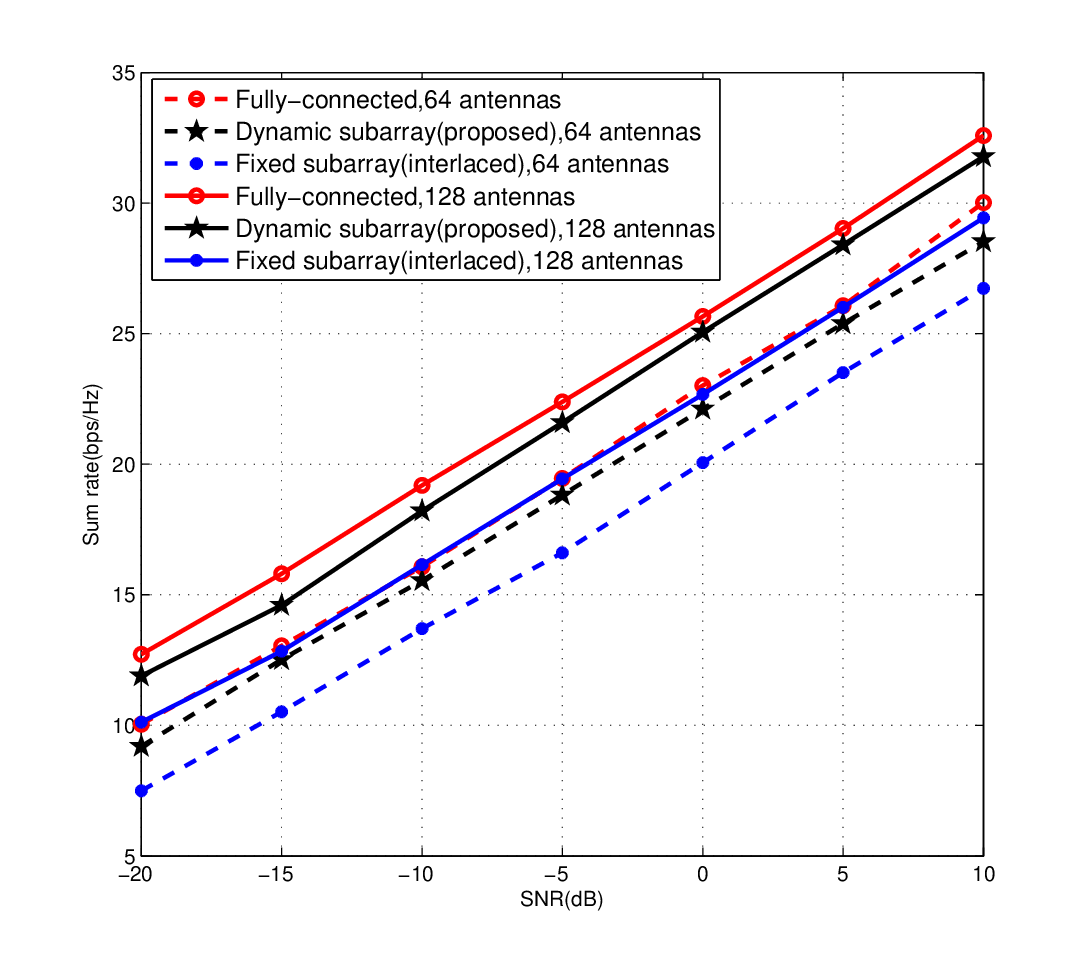}}
\subfigure[Sum rate vs. SNR (64, 128 antennas and 4 RF chains).]{
\label{fig:subfig:b} 
\includegraphics[width=3.4 in]{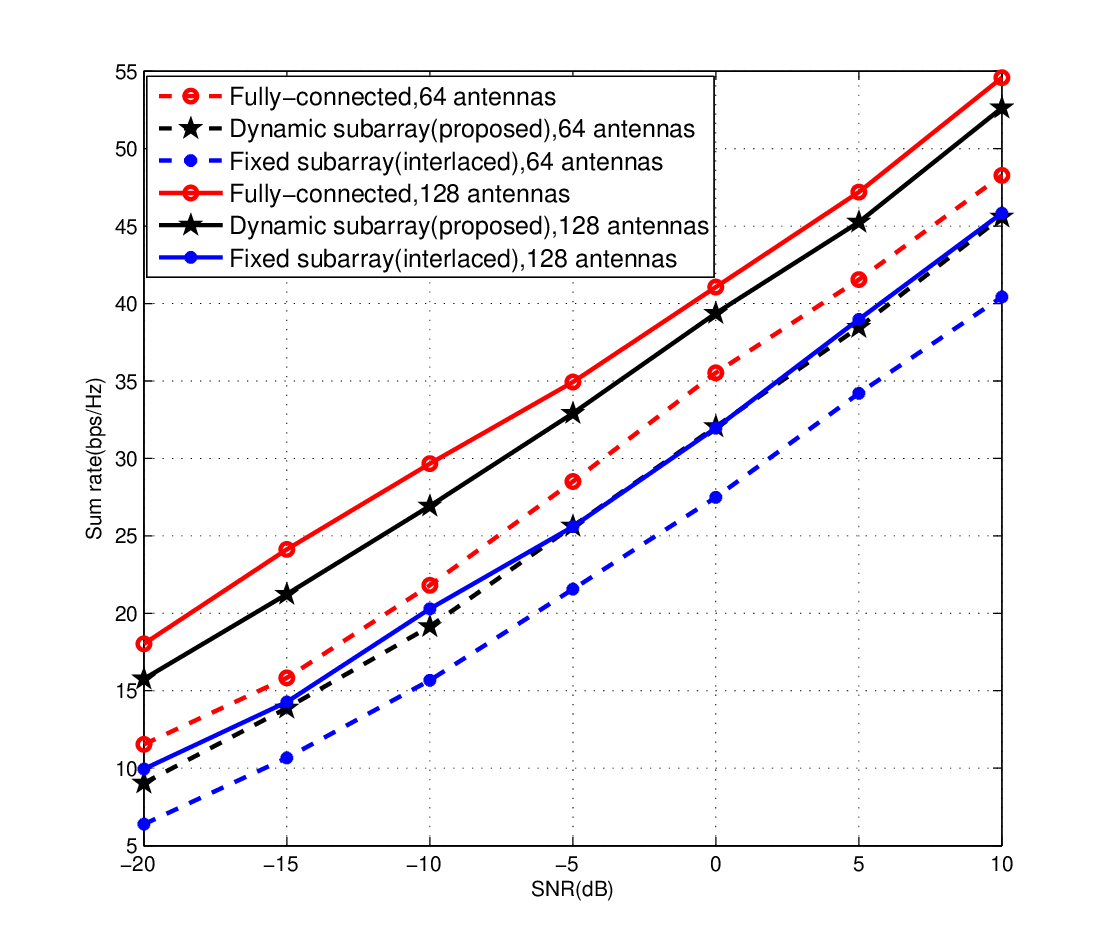}}
\hspace{0.5in}
\subfigure[Sum rate vs. SNR (64, 128 antennas and 8 RF chains).]{
\label{fig:subfig:c} 
\includegraphics[width=3.4 in]{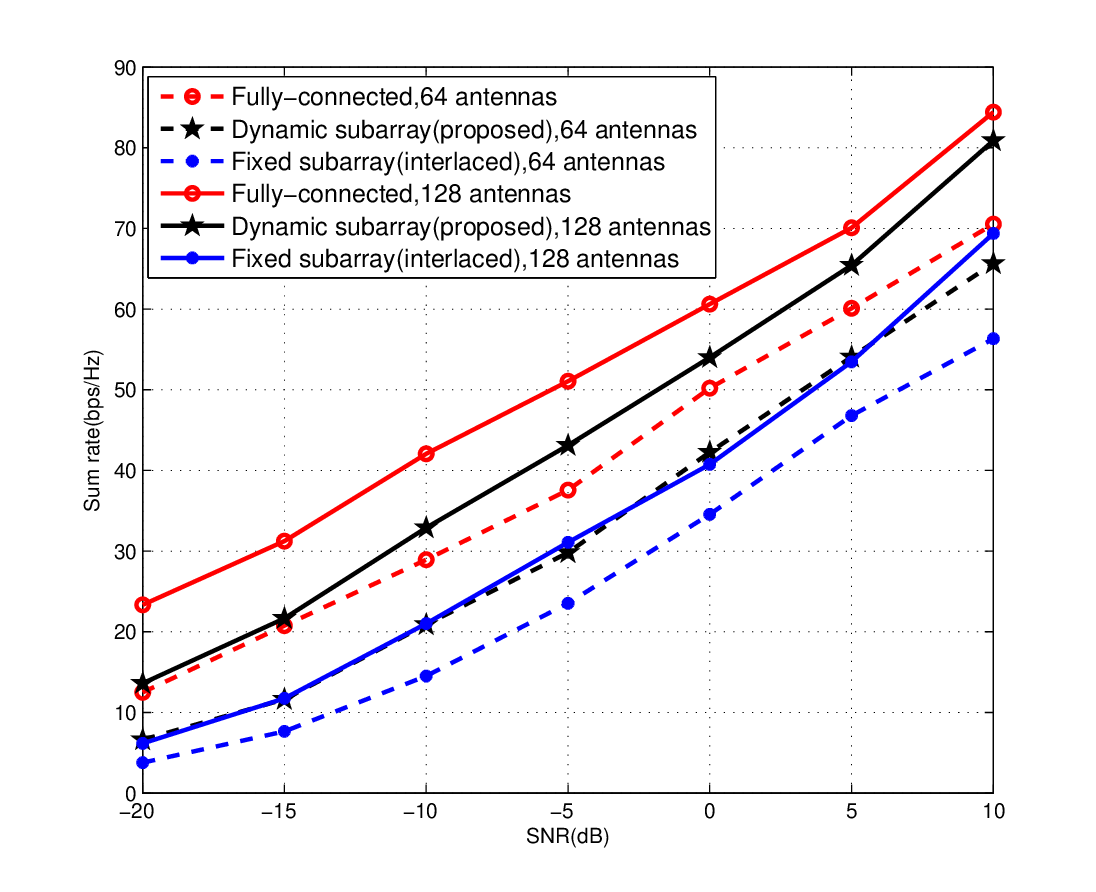}}
\caption{Performance comparisons of sum rate for different numbers of BS antennas, $N_{\rm{TX}}$.}
\end{figure}}

To further compare the performance of the proposed solution and the existing schemes, Fig. 6 indicates the sum rate according to transmit antenna numbers when each user is equipped with 2 receive antennas. Here, the numbers of RF chains at the BS are set as 2, 4, 6 and 8, respectively. In Fig. 6 (a) and (b), SNRs are assumed to be fixed at -10 dB and 0 dB, respectively. The results show that the performance of the proposed solution approaches to that of the full-connected architecture in the condition of 2 RF chains. When more RF chains are connected to the antenna arrays, the performance loss of the dynamic subarray is more obvious due to significantly reducing the hardware cost and power consumption.

Fortunately, the proposed solution can achieve a considerable high sum rate performance in the condition of $4\sim8$ RF chains and 128/256 antennas, e.g. the proposed solution can achieve about 91\% (SNR = -10 dB) and 96\% (SNR = 0 dB) of the sum rate compared with the full-connected architecture. The configurations of 128/256 antennas and $4\sim8$ RF chains at the transmitter are the most common use cases of mmWave Massive MIMO system. This result is significant for practical implementations since it means that the proposed multi-user hybrid precoding of the dynamic array performs almost as good as the full-connected architecture in the main use case of massive MIMO mmWave systems.

\textcolor{blue}{\begin{figure}[!htbp]
\centering
\subfigure[Sum rate vs. number of transmit antennas (SNR = -10 dB).]{
\label{fig:subfig:a} 
\includegraphics[width=3.6 in]{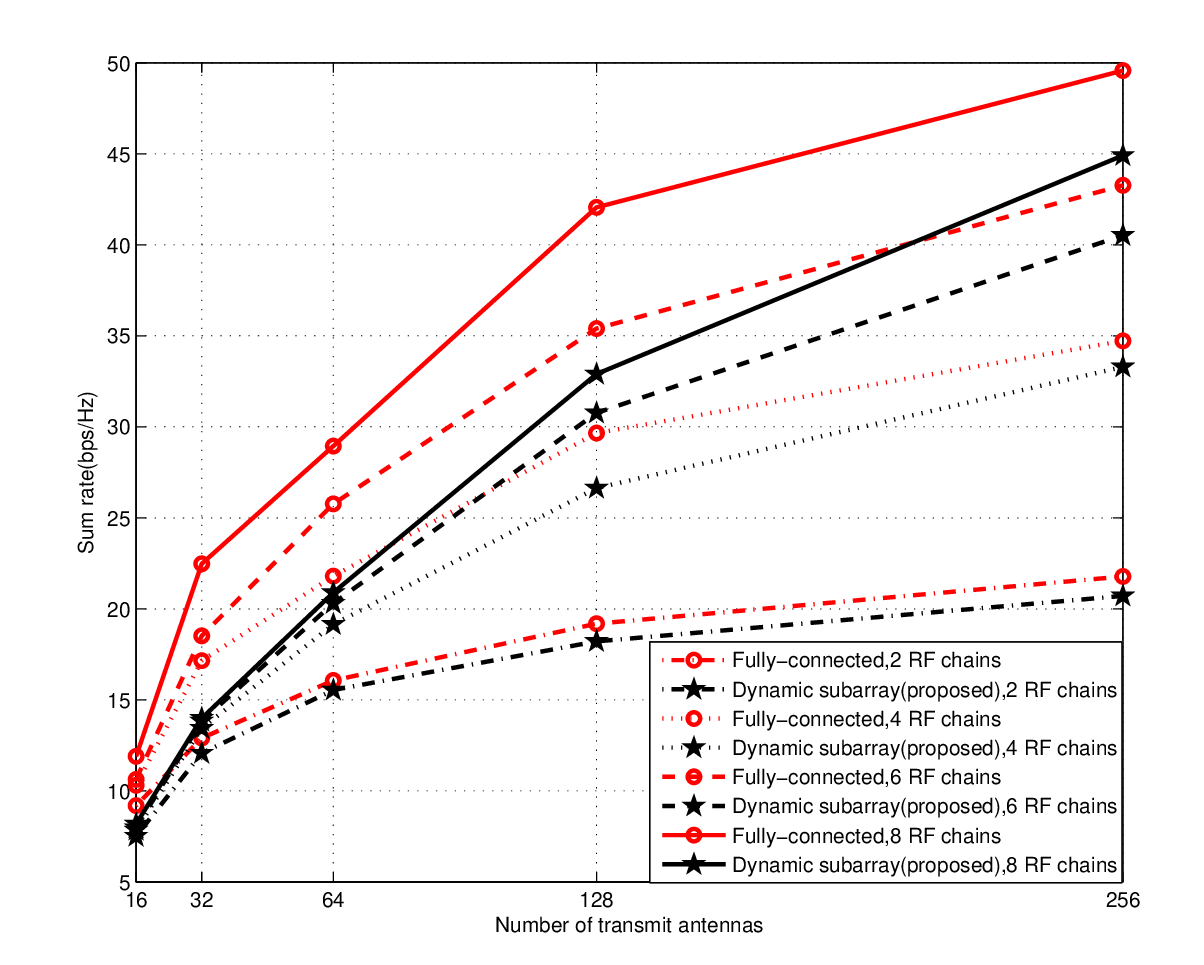}}
\subfigure[Sum rate vs. number of transmit antennas (SNR = 0 dB).]{
\label{fig:subfig:b} 
\includegraphics[width=3.7 in]{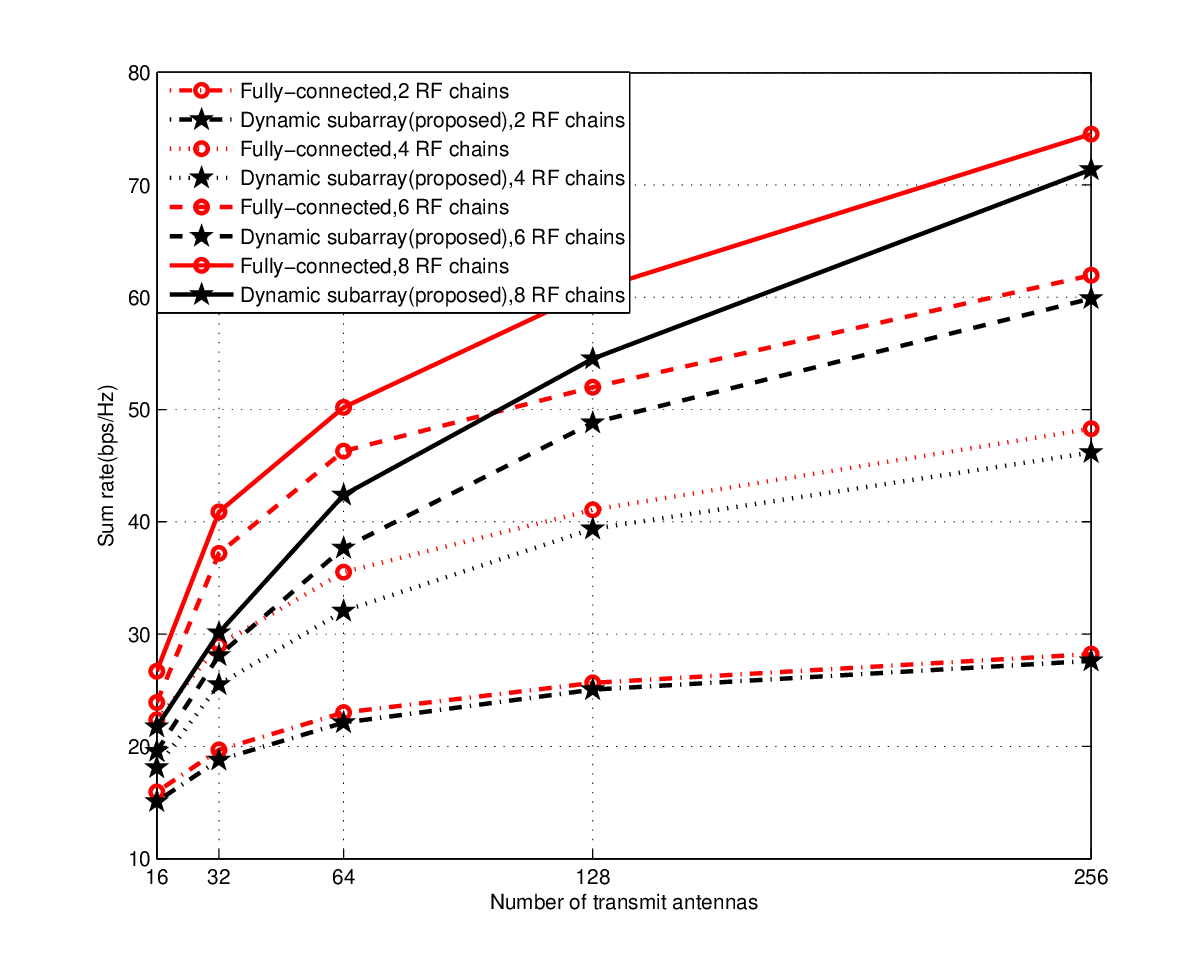}}
\caption{Comparison between 2, 4, 6 or 8 RF chains in dynamic subarray architectures with different number of antennas.}
\label{fig:subfig} 
\end{figure}}

\subsection{Performance comparisons of energy efficiency}
In this subsection, we investigate the energy efficiency of multi-user hybrid precoding designs in three antenna array architectures. According to [20], [21], the energy efficiency formula can be expressed as

\begin{align}
{\rm{EE}} = \frac{{\sum\limits_{k = 1}^K {{R_k}} }}{{{P_{total}}}} = \frac{{\sum\limits_{k = 1}^K {{R_k}} }}{{{P_t}/\eta  + {N_{\rm{RF}}}{P_{\rm{RF}}} + {N_{\rm{PS}}}{P_{\rm{PS}}}}},
\end{align} where ${P_t}$ is the transmission power constrained to $\left\| {{\bf{F}}_{\textrm{RF}}^{\mathit{k}}{\bf{F}}_{\textrm{BB}}^{\mathit{k}}} \right\|_F^2 = {N_s}$, $\eta $ is the power amplifier efficiency, ${P_{\rm{RF}}}$ is the energy consumed by RF chain, ${P_{\rm{PS}}}$ is the energy consumed by PS, and ${N_{\rm{PS}}}$ is the number of required PSs, respectively. Here, we use the ${P_{\rm{RF}}} = 250{\rm{mW}}$ [31] and ${P_{\rm{PS}}} = 1{\rm{mW}}$ [32] in simulation.

Fig. 7 shows the comparison of energy efficiency for different number of RF chains, $N_{\rm{RF}}$, where SNR = -10 dB, ${N_{\rm{TX}}} = 64$, ${N_{\rm{RX}}} = 2$. It is observed from Fig. 7 that, in accordance with the theory analysis, the increment of the number of RF chains results in performance losses on the energy efficiency evidently, since a large number of electronic equipment consumes more power. Further, we can find that all the partial-connected hybrid architectures achieve higher energy efficiency than the fully-connected hybrid architecture. Moreover, the proposed hybrid precoding solution in the dynamic sub-array obtains more energy efficient than the hybrid precoding design in the fixed subarray.

\begin{figure}
  \centering
  \includegraphics[width=3.7 in]{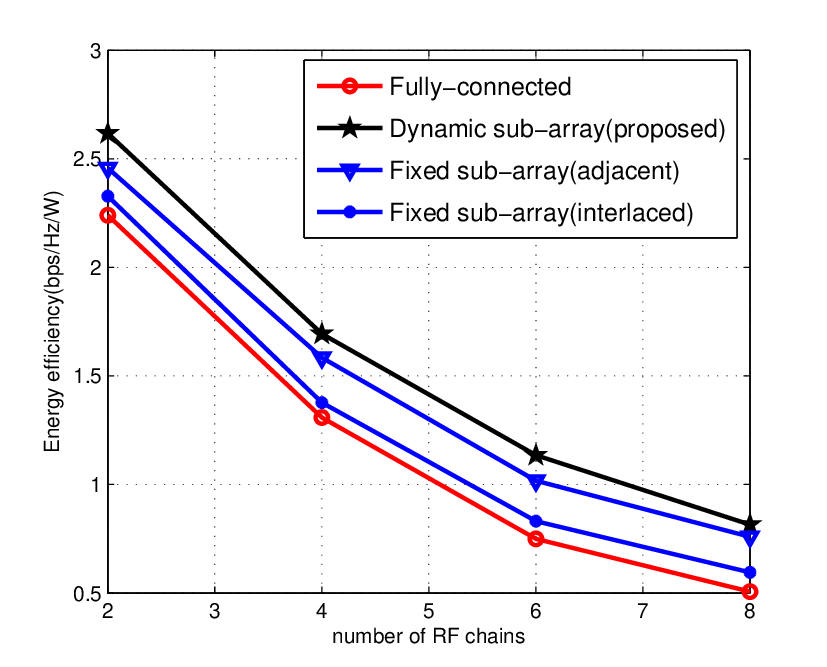}\\
  \caption{Energy efficiency comparison against the numbers of RF chains $N_{\rm{RF}}$, where ${N_{\rm{TX}}} = 64$, ${N_{\rm{RX}}} = 2$, SNR = -10 dB.}
\end{figure}

In Fig. 8, the energy efficiency versus the number of transmit antennas is illustrated with ${N_{\rm{RF}}} = 4$, ${N_{\rm{RX}}} = 2$, and SNR=-10dB. Here, BS is equipped with $N_{\rm{TX}}$ antennas, where $N_{\rm{TX}}= 16, 32, 64, 128, 256$, respectively. The results in Fig. 8 show that the improvement of energy efficiency is remarkable with the number of antennas increasing. Moreover, the proposed solution achieves better energy efficiency than the fixed subarray and the full-connected architecture.

\begin{figure}
  \centering
  \includegraphics[width=3.7 in]{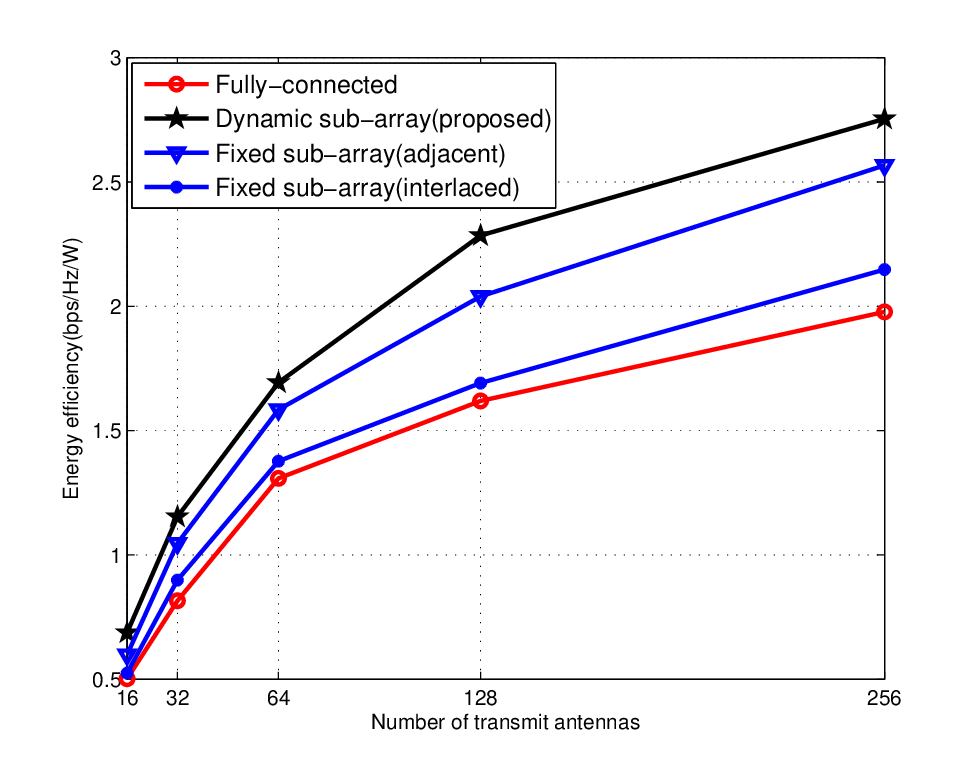}\\
  \caption{Energy efficiency comparison against the numbers of transmit antennas $N_{\rm{TX}}$, where ${N_{\rm{RF}}} = 4$, ${N_{\rm{RX}}} = 2$, SNR = -10 dB.}
\end{figure}

In conclusion, the proposed multi-user precoding for the dynamic subarray architecture obtains more energy efficient than the conventional multi-user precoding schemes according to the simulation results of energy efficiency.

\subsection{Computational complexity}
Consider a multi-user downlink system with ${N_{{\rm{RF}}}}$ RF chains and ${N_{{\rm{TX}}}}$ transmit antennas. The problem in (15) is a combinatorial optimization problem, which the exhaustive search is required to find the optimal solution from all probable cases. Such that, the computational complexity is given by

\begin{align}
\frac{1}{{({N_{{\rm{RF}}}})!}}\sum\limits_{k = 0}^{{N_{{\rm{RF}}}}} {{{( - 1)}^{{N_{{\rm{RF}}}} - k}}} \left( {\begin{array}{*{20}{c}}
{{N_{{\rm{RF}}}}}\\k\end{array}} \right){k^{{N_{{\rm{TX}}}}}}.
\end{align}

Note that, this complexity is large even for a small number of RF chains and antennas. At a consequence, the complexity of our approach can be effectively reduced compared to the conventional one by assigning each antenna. The complexity of our antenna subarray partitioning algorithm can be reduced to ${N_{{\rm{RF}}}} \times {N_{{\rm{TX}}}}$.

\section{Conclusion}
In this paper, we developed a multi-user hybrid precoding framework in the dynamic subarray of mmWave Massive MIMO systems. Especially, the proposed antenna subarray partitioning algorithm guaranteed the user fairness and reduced the computation complexity since each antenna element was allocated based on the maximal SINR increment criterion for all selected users. Simulation results showed that the sum rate and energy efficiency achieved by the dynamic subarrays architecture significantly outperforms that of the fixed subarray architectures. Furthermore, the energy efficiency of the proposed solution in the dynamic subarray case obviously outperformed that of fully-connected architecture with a slight performance loss of sum rate. Simulation results consisted with the theory analysis, and demonstrated that the proposed multi-user scheme could achieve right trade-off between the hardware complexity and system performance for mmWave Massive MIMO systems.


%

\appendices

\section*{Acknowledgment}

This paper was supported by National Natural Science Foundation of China under Grant 61871321, National Science and Technology Major Project under Grant 2016ZX03001016, and Innovation Team Project of Shaanxi Province under Grant 2017KCT-30-02.
\ifCLASSOPTIONcaptionsoff
  \newpage
\fi




\begin{thebibliography}{1}

\bibitem{IEEEhowto:kopka}
T. S. Rappaport \emph{et al.}, "Millimeter Wave Mobile Communications for 5G Cellular: It Will Work!," in \emph{IEEE Access}, vol. 1, pp. 335-349, 2013.
\bibitem{IEEEhowto:kopka}
Z. Pi and F. Khan, "An introduction to millimeter-wave mobile broadband systems," in \emph{IEEE Communications Magazine}, vol. 49, no. 6, pp. 101-107, June 2011.
\bibitem{IEEEhowto:kopka}
F. Rusek \emph{et al.}, "Scaling Up MIMO: Opportunities and Challenges with Very Large Arrays," in \emph{IEEE Signal Processing Magazine}, vol. 30, no. 1, pp. 40-60, Jan. 2013.
\bibitem{IEEEhowto:kopka}
Heath Jr., Robert W., and \emph{et al.}, "An Overview of Signal Processing Techniques for Millimeter Wave MIMO Systems," in \emph{IEEE Journal of Selected Topics in Signal Processing,} vol. 10, no. 3, pp. 436-453, Jan. 2013.
\bibitem{IEEEhowto:kopka}
O. E. Ayach, S. Rajagopal, S. Abu-Surra, Z. Pi and R. W. Heath, "Spatially Sparse Precoding in Millimeter Wave MIMO Systems," in \emph{IEEE Transactions on Wireless Communications}, vol. 13, no. 3, pp. 1499-1513, March 2014.
\bibitem{IEEEhowto:kopka}
A. Alkhateeb, O. El Ayach, G. Leus and R. W. Heath, "Channel Estimation and Hybrid Precoding for Millimeter Wave Cellular Systems," in \emph{IEEE Journal of Selected Topics in Signal Processing}, vol. 8, no. 5, pp. 831-846, Oct. 2014.
\bibitem{IEEEhowto:kopka}
C. G. Tsinos, S. Chatzinotas and B. Ottersten, "Hybrid Analog-Digital Transceiver Designs for mmWave Amplify-and-Forward Relaying Systems," \emph{2018 41st International Conference on Telecommunications and Signal Processing (TSP)}, Athens, 2018, pp. 1-6.
\bibitem{IEEEhowto:kopka}
C. G. Tsinos, S. Maleki, S. Chatzinotas and B. Ottersten, "Hybrid analog-digital transceiver designs for cognitive radio millimiter wave systems," \emph{2016 50th Asilomar Conference on Signals, Systems and Computers}, Pacific Grove, CA, 2016, pp. 1785-1789.
\bibitem{IEEEhowto:kopka}
S. Han, C. I, Z. Xu and C. Rowell, "Large-scale antenna systems with hybrid analog and digital beamforming for millimeter wave 5G," in \emph{IEEE Communications Magazine}, vol. 53, no. 1, pp. 186-194, January 2015.
\bibitem{IEEEhowto:kopka}
X. Gao, L. Dai, S. Han, C. I and R. W. Heath, "Energy-Efficient Hybrid Analog and Digital Precoding for MmWave MIMO Systems With Large Antenna Arrays," in \emph{IEEE Journal on Selected Areas in Communications}, vol. 34, no. 4, pp. 998-1009, April 2016.
\bibitem{IEEEhowto:kopka}
O. El Ayach, R. W. Heath, S. Rajagopal and Z. Pi, "Multimode Precoding in Millimeter Wave MIMO Transmitters with Multiple Antenna Sub-arrays," \emph{2013 IEEE Global Communications Conference (GLOBECOM)}, Atlanta, GA, 2013, pp. 3476-3480.
\bibitem{IEEEhowto:kopka}
Chen Y , Chen D , Jiang T , \emph{et al.} Generalized Sub-Array-Connected Hybrid Precoding Improves the Energy-Efficiency of Millimeter-Wave Massive MIMO Systems. 2018.[online].Available https://arxiv.org/pdf/1806.09246.pdf.
\bibitem{IEEEhowto:kopka}
X. Yu, J. Shen, J. Zhang and K. B. Letaief, "Alternating Minimization Algorithms for Hybrid Precoding in Millimeter Wave MIMO Systems," in \emph{IEEE Journal of Selected Topics in Signal Processing}, vol. 10, no. 3, pp. 485-500, April 2016.
\bibitem{IEEEhowto:kopka}
C. G. Tsinos, S. Maleki, S. Chatzinotas and B. Ottersten, "On the Energy-Efficiency of Hybrid Analog-Digital Transceivers for Single- and Multi-Carrier Large Antenna Array Systems," in \emph{IEEE Journal on Selected Areas in Communications}, vol. 35, no. 9, pp. 1980-1995, Sept. 2017.
\bibitem{IEEEhowto:kopka}
S. Park, A. Alkhateeb and R. W. Heath, "Dynamic Subarrays for Hybrid Precoding in Wideband mmWave MIMO Systems," in \emph{IEEE Transactions on Wireless Communications}, vol. 16, no. 5, pp. 2907-2920, May 2017.
\bibitem{IEEEhowto:kopka}
S. Park, A. Alkhateeb and R. W. Heath, "Dynamic subarray architecture for wideband hybrid precoding in millimeter wave massive MIMO systems," \emph{2016 IEEE Global Conference on Signal and Information Processing (GlobalSIP)}, Washington, DC, 2016, pp. 600-604.
\bibitem{IEEEhowto:kopka}
D. H. N. Nguyen, L. B. Le, T. Le-Ngoc and R. W. Heath, "Hybrid MMSE Precoding and Combining Designs for mmWave Multiuser Systems," in \emph{IEEE Access}, vol. 5, pp. 19167-19181, 2017.
\bibitem{IEEEhowto:kopka}
W. Ni and X. Dong, "Hybrid Block Diagonalization for Massive Multiuser MIMO Systems," in \emph{IEEE Transactions on Communications}, vol. 64, no. 1, pp. 201-211, Jan. 2016.
\bibitem{IEEEhowto:kopka}
L. Zhao, D. W. K. Ng and J. Yuan, "Multi-User Precoding and Channel Estimation for Hybrid Millimeter Wave Systems," in \emph{IEEE Journal on Selected Areas in Communications}, vol. 35, no. 7, pp. 1576-1590, July 2017.
\bibitem{IEEEhowto:kopka}
H. Lin, F. Gao, S. Jin and G. Y. Li, "A New View of Multi-User Hybrid Massive MIMO: Non-Orthogonal Angle Division Multiple Access," in \emph{IEEE Journal on Selected Areas in Communications}, vol. 35, no. 10, pp. 2268-2280, Oct. 2017.
\bibitem{IEEEhowto:kopka}
Q. Xue, X. Fang, M. Xiao and L. Yan, "Multiuser Millimeter Wave Communications With Non-orthogonal Beams," in \emph{IEEE Transactions on Vehicular Technology}, vol. 66, no. 7, pp. 5675-5688, July 2017.
\bibitem{IEEEhowto:kopka}
A. Li and C. Masouros, "Hybrid Analog-Digital Millimeter-Wave MU-MIMO Transmission With Virtual Path Selection," in \emph{IEEE Communications Letters}, vol. 21, no. 2, pp. 438-441, Feb. 2017.
\bibitem{IEEEhowto:kopka}
X. Zhu, Z. Wang, L. Dai and Q. Wang, "Adaptive Hybrid Precoding for Multiuser Massive MIMO," in \emph{IEEE Communications Letters}, vol. 20, no. 4, pp. 776-779, April 2016.
\bibitem{IEEEhowto:kopka}
M. R. Akdeniz \emph{et al.}, "Millimeter Wave Channel Modeling and Cellular Capacity Evaluation," in \emph{IEEE Journal on Selected Areas in Communications}, vol. 32, no. 6, pp. 1164-1179, June 2014.
\bibitem{IEEEhowto:kopka}
P. Kyosit \emph{et al.} IST-4-027756 WINNER II D1.1.2 V1.2 WINNER II Channel Models. [Online]. Available: www.ist-winner.org.
\bibitem{IEEEhowto:kopka}
D. P. Palomar, J. M. Cioffi and M. A. Lagunas, "Joint Tx-Rx beamforming design for multicarrier MIMO channels: a unified framework for convex optimization," in \emph{IEEE Transactions on Signal Processing}, vol. 51, no. 9, pp. 2381-2401, Sept. 2003.
\bibitem{IEEEhowto:kopka}
D. P. Palomar and Mung Chiang, "A tutorial on decomposition methods for network utility maximization," in \emph{IEEE Journal on Selected Areas in Communications}, vol. 24, no. 8, pp. 1439-1451, Aug. 2006.
\bibitem{IEEEhowto:kopka}
T. E. Bogale, L. B. Le, A. Haghighat and L. Vandendorpe, "On the Number of RF Chains and Phase Shifters, and Scheduling Design With Hybrid Analog-Digital Beamforming," in \emph{IEEE Transactions on Wireless Communications}, vol. 15, no. 5, pp. 3311-3326, May 2016.
\bibitem{IEEEhowto:kopka}
Taesang Yoo and A. Goldsmith, "On the optimality of multiantenna broadcast scheduling using zero-forcing beamforming," in \emph{IEEE Journal on Selected Areas in Communications}, vol. 24, no. 3, pp. 528-541, March 2006.
\bibitem{IEEEhowto:kopka}
M. Schubert and H. Boche, "Solution of the multiuser downlink beamforming problem with individual SINR constraints," in \emph{IEEE Transactions on Vehicular Technology}, vol. 53, no. 1, pp. 18-28, Jan. 2004.
\bibitem{IEEEhowto:kopka}
P. Amadori and C. Masouros, "Low RF-complexity millimeter-wave beam space-MIMO systems by beam selection," IEEE Trans. Commun., vol. 63, no. 6, pp. 2212-2222, Jun. 2015.
\bibitem{IEEEhowto:kopka}
C. A. Balanis, Antenna Theory: Analysis and Design. Hoboken, NJ, USA: Wiley, 2012.
\end{thebibliography}
%

%
\begin{IEEEbiography}[{\includegraphics{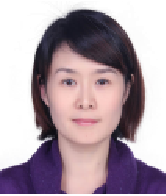}}]{Jing Jiang}
received M. Sc. from the Xi Dian University in 2005 and a Ph.D. in Information and Communication Engineering from North Western Polytechnic University, China in 2009. She had been a researcher and a project manager from 2006 to 2012 at ZTE Corporation in China, and currently as a professor of Shaanxi Key Laboratory of Information Communication Network and Security, Xi'an University of Posts and Telecommunications, Xi'an, China. Her research interests include massive multiple-input multiple-output systems and millimeter-wave communications. She has been a Member of the IEEE, 3GPP.
\end{IEEEbiography}

\begin{IEEEbiography}[{\includegraphics{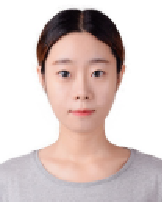}}]{Yue Yuan}
 received the B.S. degree in Xi'an University of Posts and Telecommunications, Xi'an, China, in 2015. She is currently pursuing the M.S. degree in Xi'an University of Posts and Telecommunications, Xi'an, China. Her research interests include the massive multiple-input multiple-output systems and 5G millimeter-wave communications. She is a member of the Shaanxi Key Laboratory of Information Communication Network and Security, Xi'an University of Posts and Telecommunications.

\end{IEEEbiography}

\begin{IEEEbiography}[{\includegraphics{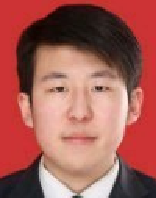}}]{Li Zhen}
 received the B.S. degree in electronic information science and technology from the Xi'an University of Science and Technology, Xi'an, China, in 2009, and the M.S. degree in circuit and system from Xi'an University of Posts and Telecommunications, Xi'an, China, in 2012, and the Ph.D. degree in communication and information systems from Xidian University, Xi'an, China, in 2018. He is currently a lecturer with the Xi'an University of Posts and Telecommunications, Xi'an, China. His research interests include weak signal detection and processing, random access, and 4G/5G-satellite mobile communications.
\end{IEEEbiography}



\end{document}